\documentstyle[sprocl,epsfig,axocolor]{article}

\def\Journal#1#2#3#4{{#1} {\bf #2}, #3 (#4)}

\def\NIM{\em Nucl. Instrum. Methods}

\def\NPB{{\em Nucl. Phys.} B}
\def\PLB{{\em Phys. Lett.}  B}
\def\PLA{{\em Phys. Lett.}  A}
\def\PRL{\em Phys. Rev. Lett.}
\def\PRD{{\em Phys. Rev.} D}
\def\ZPC{{\em Z. Phys.} C}
\def\ZP{\em Z. Phys.}
\def\PR{\em Phys. Rev.} 

\def\be{\begin{equation}}
\def\ee{\end{equation}}
\def\bea{\begin{eqnarray}}
\def\eea{\end{eqnarray}}
\def\eplem{\mbox{$e^+e^-$}}
\def\ptmin{\mbox{$p_{Tmin}$}}
\def\gamgam{\mbox{$\gamma \gamma$}}
\def\siginel{\mbox{$\sigma^{inel}_{\gamma \gamma}$}}
\def\sigjet{\mbox{$\sigma^{\rm jet}_{e^+e^- \gamma \gamma}$}}
\def\siggmjet{\mbox{$\sigma (\gamma \gamma \rightarrow {\rm jets})$}}
\def\twophoton{\mbox{$2 \gamma$}}
\def\fpgone{\mbox{$f_{P_1/\gamma_1}(x_1)$}}
\def\fpgtwo{\mbox{$f_{P_2/\gamma_2}(x_2)$}}
\def\pt{\mbox{$p_T$}}
\def\kt{\mbox{$k_T$}}
\def\Phad{\mbox{$P_{had}$}}     
\def\RedViolet{} 
\def\Black{}

\begin{document}
\begin{flushright}
hep-ph/9807379 \\
IISc-CTS/8/98
\end{flushright}
\title{BEAMSTRAHLUNG INDUCED MINIJET BACKGROUNDS AT FUTURE COLLIDERS
AND A MODEL FOR $\sigma^{inel}_{\gamma \gamma}$\footnote{Invited talk 
presented at the International Confernce on Quantum Aspects of 
Beam Physics, Jan. 1998, Monterey, U.S.A.}}

\author{R.M. GODBOLE\footnote{On leave of absence from University
of Bombay, Mumbai, India.}}

\address{Center for Theoretical Studies, Indian Institute of Science,
\\ Bangalore, 560 012, India,\\E-mail: rohini@cts.iisc.ernet.in}


\maketitle\abstracts{The phenomenon of beamstrahlung can give 
rise to potentially dangerous hadronic backgrounds due to
minjet production at the future linear colliders as well
as at the $\gamma \gamma$ colliders that are under consideration.
In this talk I will review briefly the current estimates of these
backgrounds and predictions of the eikonalised minijet model
for \siginel. I end by pointing out issues that need to be studied 
in more detail to firm up our estimates of these backgrounds.}

\section{Introduction}
The phenomenon of beamstrahlung~\cite{beam} at high bunch densities 
is one of the most interesting quantum aspect of the beam physics 
(QABP). I want to begin this discussion by pointing out the particle 
physics facts which make it clear that in the quest of ever rising
energies,  the phenomenon of beamstrahlung is unavoidable 
in the energy range ($300 \leq \sqrt s \leq 2000 $ GeV)
of the next generation \eplem\ colliders (NLC's).  I then 
want to summarise 
briefly how photons develop `strong' interactions at high energies and
how these combined with beamstrahlung photons can give rise to a new class
of hadronic backgrounds~\cite{prlus,zpcus,chen} at the NLC's. 
Since these backgrounds are caused by the `hadronic' interactions of the
photon, they are relevant (perhaps even more so) for the
\gamgam\ colliders that are being planned using backscattered 
laser photons. I will then comment on the current estimates of 
these backgrounds clearly pointing out the sources of 
uncertainties. I will then  present a newer, convenient parametrisation
of the \gamgam\ minijet cross-sections incorporating the recent 
information on the photon structure function as well as on \siginel. 
I will also present results of a new calculation of \siginel\ in an 
eikonalised minijet model \cite{liame} and  will end by pointing out the 
improvements necessary in the estimates of the minijet induced 
backgrounds.

\section{Particle Physics and beamstrahlung at NLC's}

The 'clean' environment of the \eplem\ colliders has played an 
essential role in the developement of particle physics. A quick 
comparison of the cross-sections of physically interesting
processes such as $e^+e^- \rightarrow f \bar f$, 
($e^+e^- \rightarrow W^+W^-$), at the current \eplem\ colliders 
like LEP-1(LEP 200) ($\sim 1000 (30)$ pb) to the ones expected
at the NLC's ($\sim 1-10$ pb), shows that the NLC's will have to
operate at luminosities at least 3-4 orders of magnitude higher than
those at LEP. This has to be coupled with the fact that due to the
higher energies these colliders will have to be {\it linear}
colliders. If we recall that at a storage ring collider like LEP, a given
bunch circulates $\sim 10^8$ times, it is clear that the bunch densities
required at the NLC's will have to be very high indeed. It can be seen
that for almost all the designs under consideration there will be
beamstrahlung, the spectra~\cite{zpcus,chen,hawaiius} depending
on the machine designs. A very convenient parametrisation for the calculation
of the beamstrahlung spectra, for small values of the beamstrahlung parameter
$\Upsilon$ has been given by Chen~\cite{chenan}. Thus along with each
\eplem\ collision there will also be an underlying \gamgam\ collision
due to the beamstrahlung photons. 

If  the construction of \gamgam\ colliders using  backscattered laser 
beams~\cite{telnov} becomes a reality, which might be possible as indicated 
by recent experiments from SLAC~\cite{slacprl}, the luminosity and 
the energy of the
\gamgam\ collisions will be close to those of the parent \eplem\
machine. Just like the parent NLC's the \gamgam\ colliders may 
also suffer from the minijet induced backgrounds due to the photon
structure.

\section{Minijet induced backgrounds due to hadronic structure of photon}
\subsection{Hadronic structure of photons and `resolved' contributions
to jet cross-section in \protect\twophoton\  processes. \label{sec:res}}

Before beginning to discuss the size of the backgrounds caused by the
combination of the hadronic structure of the photon and the phenomenon of
beamstrahlung at the NLC's, let us briefly recapitulate what is 
known experimentally about \gamgam\ interactions at present.
All the current  information comes from the study of jet/hadron
production in the two photon processes
\be
e^+ e^- \rightarrow e^+e^- \gamma \gamma \rightarrow {\protect \rm hadrons}
 + e^+ e^-
\label{twogam}
\ee
where the $e^+,e^-$ are lost along the beam pipe. In this process the
photons are `almost' real and the spectrum is ordinary 
Weisz\"acker-Williams (WW)~\cite{ww} spectrum. The $2 \gamma$ physics
was studied for the first time at PEP/PETRA with $\sqrt s 
= W_{\gamma \gamma} = 5-10 $ GeV. 
Strong interactions of the photon arise due to its fluctuations into a 
short  lived $\bar q q$ pair. At high energies and hence on shorter time 
scales, this can be computed using perturbative QCD \cite{witten}. The
photon thus can be looked upon as surrounded by a `cloud' of quarks and
gluons. Some part of the interactions of the high energy photons can
therefore be computed as though the photon `consists' of these partons.
The early theoretical investigations
\cite{tristanus} indicated that already at TRISTAN the jet production 
in \twophoton\ interactions would be dominated by the
`resolved' processes \cite{reviewus}, where the partonic constituents 
of the photon and not the photon itself, participate in the hard scattering
process giving rise to jets. 

There exist  three different types of contributions to jet production
in \gamgam\ processes where both/one/no photon take part `directly'
in the hard process. We call these direct/1-resolved/2-resolved processes
respectively. Schematically one can write down the jet production 
cross-section as follows:
\bea
{d \sigma \over d \pt} \Big|^{jet}_{2-res}& =& \int dz_2 \int dz_1
\int dx_1 \int dx_2 ~~f_{\gamma/e^+} (z_1) f_{\gamma/e^-} (z_2)\nonumber \\ 
&&{}\sum_{P_1,P_2,P_3,P_4} \fpgone \fpgtwo \;\; 
{d \sigma \over d\pt} (P_1 P_2 \rightarrow P_3 P_4).
\label{schem}
\eea
Here $P_i (i=1,4)$ denote the partons, $f_{a/b} (x)$ denotes the flux of
partons of type `a' in `b'  carrying momentum fraction `x' of `b'.  For 
the `1-res' (`direct') processes one (both) the flux factors 
$f_{P_i/\gamma}(x_i)$ are to be replaced by $\delta (1-x_i)$, 
corresponding to the fact that the entire energy of the 
photon is available for the scattering process giving rise to the jet
production. For the case of TRISTAN/LEP1/LEP 200 etc. the photon
flux factors $f_{\gamma/e} (z)$  are just the WW flux factors, whereas
the beamstrahulng flux will have to be added to these at the NLC's.

The \twophoton\ processes in general dominate production of
low invariant mass hadron production over the annihilation
processes at higher energy, due to the logarithmic enhancement
of the WW flux factors.  Our investigations \cite{tristanus} indicated 
that  the jet production in \twophoton\ processes is dominated by `resolved'
processes already for TRISTAN energies; {\it i.e.}

\be
{d \sigma \over d \pt} \Big|^{jet}_{direct} <
{d \sigma \over d \pt} \Big|^{jet}_{1-res} <
{d \sigma \over d \pt} \Big|^{jet}_{2-res} 
\label{order}
\ee
The existence of resolved contributions in  \twophoton\ production of
jets was clearly demonstrated by the TRISTAN experiments \cite{tristandta}.
The dominance of the jet production by the `resolved' processes 
increases with energy. This is demonstrated in 
\begin{figure}[htb]
\leavevmode
\begin{center}
\mbox{\epsfig{file=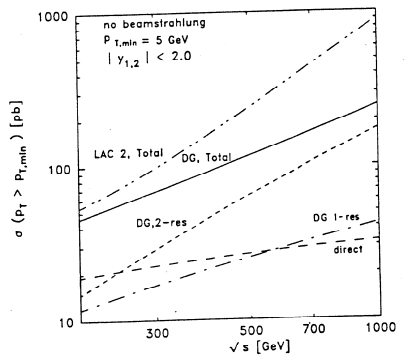,height=5cm}}
\end{center}
\caption{$p_T$  integrated, central jet production cros-section in {\protect 
\twophoton} processes as a function of $\sqrt s$. \label{fig1}} 
\end{figure}
Fig. \ref{fig1}. The figure shows  cross-section for production of central jets
($|y| < 2$) integrated above a certain \ptmin\ in \twophoton\ processes
in \eplem\ collisions, as a function of $\sqrt s$. Here we notice the 
following,
\begin{itemize}
\item If we recall that at these energies $\sigma (e^+e^- \rightarrow W^+ W^-)$
is $\simeq 10 $ pb, we realise that the cross-section for {\it inclusive} 
production of central jets in \twophoton\ processes at the NLC's
is very large indeed.
\item It is dominated by the `resolved' processes. 
\item It also has a strong dependence on the assumed parton content of
the photon. The figure shows the results for two parametrisations of
the photonic parton densities \cite{param}. 
\item The cross-section increases almost linearly with $\sqrt s$. Of course
this has no problem with unitarity as this is an {\it inclusive} cross-section.
\end{itemize}
Fig. \ref{fig1} above shows the expected cross-sections {\it without }
inclusion of any beamstrahlung photons.  Inclusion of the beamstrahlung
contribution to the spectra show that  the predicted cross-sections vary 
widely, depending on the machine design even for a given $\sqrt s$. 
From the point of view of
the minjet-induced hadronic backgrounds more interesting is the size of the
\sigjet\ cross-sections for lower values of \ptmin. We show in 
\begin{figure}[htb]
\leavevmode
\begin{center}
\mbox{\epsfig{file=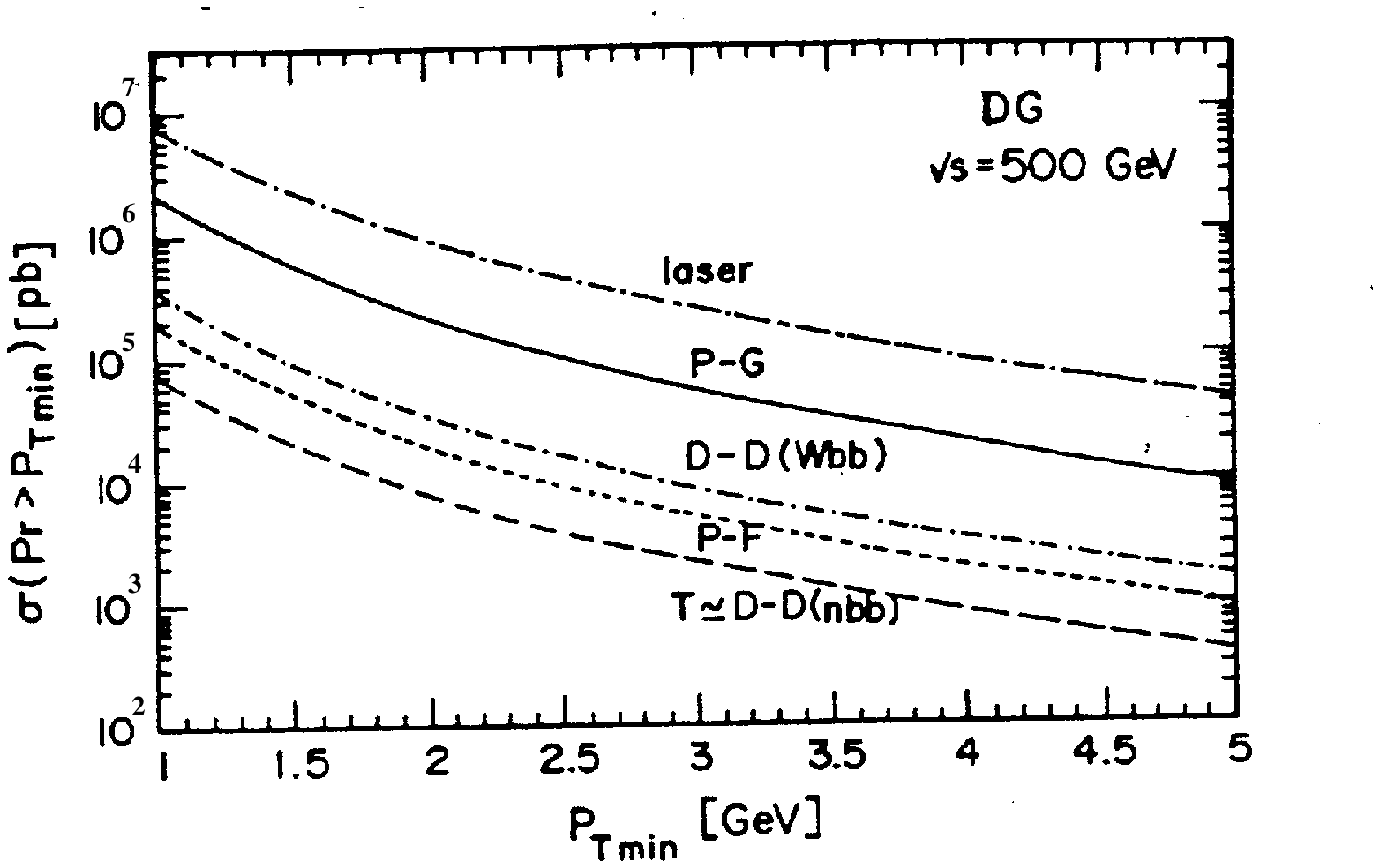,height=5.5cm}}
\end{center}
\caption{$p_T$  integrated jet production cros-section in {\protect 
\twophoton} processes as a function of \protect\ptmin for different
machine designs with \protect $\sqrt s$ = 500 GeV including the
beamstrahlung photons. \label{fig2}} 
\end{figure}
Fig.~\ref{fig2} the effect of including beamstrahlung contribution, at a fixed
$\sqrt s$. The contribution {\it without} any beamstrahlung, though not shown in
this figure is pretty close to the curve labelled T, as for this design (TESLA)
the beamstrahlung spectrum is pretty soft. We notice from Fig.~\ref{fig2} the following:
\begin{itemize}
\item Inclusion of beamstrahlung increases
$\sigjet \simeq \sigma (\eplem \rightarrow \eplem \gamgam \rightarrow$ 
$\eplem + {\rm jets}\;)\big|_{\pt > \ptmin}$, 
by an order of magnitude or more as can be seen by comparing 
Fig.~\protect\ref{fig1} and Fig.~\protect\ref{fig2}.
\item The different legends on the figure refer to different machine
designs (for details see \protect\cite{zpcus}). We can see that the predicted 
jet cross-sections  depend on the machine designs in a striking manner, 
the dependence of course reflecting the different hardness of the beamstrahlung
spectra expected. There is no cut on the rapidities of jets here.
\item The value of \sigjet\ of course depends very strongly 
on \ptmin\ and on the parton content of the photon. For the parton 
distributions used in the figure here (DG~\cite{DG}) $\sigjet\ \sim \ptmin^{(-3.3)}$.
\item The curve labelled {\bf Laser} is the one expected for a \protect\gamgam\
collider constructed using the spectra as given in Ref. [8].
\item Note that here we have decreased \protect\ptmin\ to a 
very low value ($\ge 1$ GeV). This means we have strictly gone to the limit of the applicability
of perturbation theory. We will discuss the implication of this for the 
minijet induced backgrounds in the next subsection.
\end{itemize}
\subsection{Minijet induced backgrounds at NLC's}
The discussion of Sec.~\ref{sec:res} shows that the \sigjet\
does indeed grow very significantly  at the high energy 
\eplem\  colliders with energy with the inclusion of beamstrahlung 
and at high energy \gamgam\ colliders. The  results of 
Fig.~\ref{fig2} show that the `minijet' cross-sections do indeed rise very fast and are very large at the NLC's. This indicates that associated with every 
`effective' bunch crossing there will  be underlying 
events with low \pt\ hadrons being produced from the \gamgam\ 
collision which have nothing to do with the \eplem\ event under study. 
In our original study we constructed the
following measure for the `messiness' of an event. We first defined an
effective luminosity per bunch crossing in the following fashion. If  
$\delta t$ is the time resolution of the detector , $\Delta t$ 
is the temporal separation between different bunches and ${\cal L}_b$ is
the luminosity per bunch crossing, 
\bea
{\cal L}_{eff} & =&{\cal L}_b , \;\;\;\;\;\;\; {\em if}\;\;  \Delta t > \delta t
\nonumber\\
& =& {\cal L}_b \times {\delta t \over \Delta t}, \;\;\; {\em if}\;\; \Delta t < \delta t
\eea
The `effective' number of `minijet' events which will then produce an 
udnerlying event will be given by multiplying the cross-sections given in
Fig.~\ref{fig2} by ${\cal L}_{eff}$ for different machine designs. In
\begin{table}[htb]
\begin{center}
\caption{Estimate of minijet-induced `messiness' for different
designs of NLC's \label{messiness}}
\vspace{0.2cm}
\footnotesize  
\begin{tabular}{|c|c|c|c|}
\hline
&&&\\
{Collider} & {$\sigma^{Semi-hard}$ $(\mu $b)} & {$\sigma^{soft}$ ($\mu$b)} &No. of minijet events\\ 
&&&\\
\hline
&&&\\
    T      &        0.016           &         0.041              &   0.004 \\
&&&\\
 D-D (nbb) &        0.020           &         0.051              &   0.021  \\
 &&&\\ 
 P-F       &        0.042           &         0.072              &   0.46    \\
 &&&\\
 JLC1      &        0.069           &         0.12               &    1.1    \\
&&&\\
\RedViolet
$\gamma \gamma $ (500 )& 1.9        &         0.25               & $>\sim O(10-20)$\\
&&&\\
\Black
T(1000)    &        0.057           &        0.099               & 0.0036 \\
&&&\\
T(2000)    &        0.21            &        0.13                & 0.013  \\
&&&\\
$T^\prime (1000)$ & 0.17            &        0.27                & 0.043  \\
\hline
\end{tabular}
\end{center}
\end{table}     
Table~\ref{messiness} taken from Ref.~[3], I show the number 
of `minijet' events expected per effective bunch crossing. If this number 
is well below 1 then the collider has no significant minijet-induced, 
beamstrahlung dependent hadronic backgrounds. In Table~\ref{messiness}, 
$\sigma^{\em Semi-hard}$ is the 
above mentioned `minijet' cross-section. The value of \ptmin\ used in this
Table is 1.6 GeV. As seen from Fig.~\ref{fig2},  \sigjet\ depends strongly 
on the value of \ptmin . Hence that is indeed a major source of uncertainty
of the minijet-induced messiness. We will comment on it a little later.
\sigjet\ calculated above can basically be used as a figure of merit for 
different machines. Table~\ref{messiness} shows that by choosing the machine
design judiciously and reducing beamstrahlung, it is possible to reduce the 
minijet-induced hadronic background to an acceptable level for \eplem\
colliders upto a $\sqrt s$ of $\sim 1000 $ GeV. However, for a `Laser' collider
even for $\sqrt s = 500 $ GeV, the expected number of `minijet' events 
is unacceptably large. Also please note that in calculating the last column
we have not inlcuded the contribution from $\sigma^{\em soft}$ as most of 
the hadrons produced in the soft \gamgam\ reactions will be lost along the
beam pipe. This contribution is essentially computed using a constant 
$\gamgam \rightarrow {\em hadrons}$ cross-section and using $W_{\gamgam} > 
5$ GeV. 

Here a comparison with the results~\cite{chen} obtained by Chen 
and collaborators  is in order. They, very correctly, stressed that 
while it is clear that $\sigjet $ does grow in the manner we noticed, 
the  following three questions are also relevant in assessing this
background: 1) How much of the increase of $\siggmjet$ is reflected in 
the increase in $\siginel $, 2) How many of the hadrons produced 
by these minijet interactions reach the detectors; i.e. the issue
of the scalar $E_T$  distribution and 3) How best to determine the value of 
\ptmin. At the end of their analysis they also conclude that it is possible to
design \eplem\ colliders upto $\sqrt s \sim 1$ TeV which are free from
minijet backgrounds. It is worth pointing out here that the
`minijet crisis' here has vanished not because the estimates of backgrounds
have been lowered but the beam designs have been  modified. 
Recently (see talks at these conferences) people have started  thinking about yet higher energy \eplem\ colliders, with $\sqrt s \leq 5 $ TeV. I will comment
about the estimates of the minijet backgrounds at these in the last
section.  We do however, disagree with Chen et al, in our onclusion 
for $\gamgam $ colliders. The two analyses did differ greatly in 
the choice of \ptmin\ and we will comment on it in the next section. 

It is true that for \gamgam\ colliders effects of multiple parton interactions
and hence eikonalisation becomes more important than in the case of
\eplem\ colliders. Since  the analysis of Ref.~[4],  full MC
generators for \gamgam\ events at TRISTAN/LEP, including the
`resolved' contributions have become available and 
for a better assesment of the `minijet' background, perhaps a newer 
MC analysis is needed. Since the size of these backgrounds depend crucially
on the parton densities in photon, it is important to also include the updated
parametrisations~\cite{GRV,SAS} that are available now. A 
determination of \ptmin\ using the latest data~\cite{LEPL3,OPAL} on \siginel\
is also necessary. In the analysis of Ref. [4] the \siggmjet\ was eikonalised. 
In the eikonalisation it was assumed that the multiple parton interaction
causing it are completely independent of each other. The existence
of `pedestal' effect shows that that is perhaps not completely correct.

In general the issue of how much of the rise of \siggmjet\
is reflected in \siginel\ is an important one. While I discuss that in the 
next section, here I give a newer parametrisation of the `minijet' 
cross-sections in  \gamgam\ collisions which can be used in estimating the
hadronic backgrounds at the NLC's by folding it with appropriate beamstrahlung
spectra. This supercedes the corresponding parametrisation that was given 
in~\cite{zpcus}.

Our discussions of the next section will show that for \gamgam\ interactions 
$\ptmin = 2$ GeV is a choice consistent with the data on \siginel\ 
in the context of an eikonalised minijet model. The cross-section  
\be
\sigma_{minijet}\equiv
\sigma (\gamma \gamma \rightarrow {\rm jets})\big|_{\ptmin}^{\sqrt s} \equiv
\int_{\ptmin} 
{d \sigma \over d \pt} (\gamgam \rightarrow {\rm jets})
\label{Sigjet}
\ee 
for the two parametrisations GRV~\cite{GRV} and SAS~\cite{SAS} densities  
is given (in nb) 
\bea 
\sigma_{minijet}& = &
\left[222 \left({2 \;{\rm GeV} \over \ptmin}\right)^2 - 161 \left({2\; {\rm GeV} \over \ptmin}\right) 
+ 36.6 \right] \left({\sqrt s\over 50}\right)^{1.23} \label{grv}\\
& =&\left[77.6 \left({2\; {\rm GeV} \over \ptmin}\right)^2 - 45.9 
\left({2\; {\rm GeV} \over \ptmin}\right) + 9.5 \right] \left({\sqrt s \over 50}\right)^{1.17}
\label{eq:sas}
\eea
by Eqs.~\ref{grv} and \ref{eq:sas} respectively. Here $\sqrt s$ is in GeV.
Since the dependence on \ptmin\  of \siggmjet\ is extremely strong,
it is essential to fix that  well.
One way of fixing the correct choice of \ptmin\ is to use the eikonalised
minijet model to describe the observed features of the \twophoton\ reactions
at TRISTAN/LEP. To that end in the next section I present a new calculation
of \siginel\ in the eikonalised minijet model.

I end this section by commenting on the new theoretical issues that will 
have to be taken in to account in extending these calculations to the higher
energy ($\sqrt s \leq 3-5$ TeV) \eplem\ and \gamgam\ colliders. At these
energies the $x_{\gamma}$ values at which photonic parton densities will 
be sampled will be small ($\simeq 10^{-3}$) and hence saturation effects 
might have to be taken into account. At present, no complete
theoretical discussion of the subject is available.

\section{An eikonalised minijet model for \siginel\ }
As is well known, even though the `minijet' cross-sections rise very fast
with energy (as shown by Eqs.~\ref{grv} and \ref{eq:sas}) the \siginel\ 
certainly do not rise that fast. The `minijet' cross-section has to be 
eikonalised so that unitarity is not violated. In general for photon 
induced processes, the inelastic cross-section obtained by eiknolisation
(and hence unitarisation) of the minijet cross-section is given by
\begin{equation}
\label{eikonal}
\sigma^{inel}_{ab} = P^{had}_{ab}\int d^2\vec{b}[1-e^{n(b,s)}]
\end{equation}
with the average number of collisions at a given impact
parameter $\vec{b}$ given by  
\begin{equation}
\label{av_n}
n(b,s)=A_{ab} (b) (\sigma^{soft}_{ab} + {{1}\over{P^{had}_{ab}}}
\sigma^{jet}_{ab})
\end{equation} 
where $P^{had}_{ab}$ is the probability that  the colliding particles
$a,b$ are both in a hadronic state,  
$A_{ab} (b)$ describes the transverse overlap of the  partons 
in the two projectiles  normalised to 1,
$\sigma^{soft}_{ab}$ is the non-perturbative part of the cross-section
while $\sigma^{jet}_{ab} $ is the hard part of the cross--section (of order
$\alpha $ or $\alpha^2$ for $\gamma p$ and $\gamma \gamma$ respectively). 
Notice that, in the above definitions, $\sigma_{soft}$
is a cross-section of hadronic size since the factor $P_{ab}^{had}$ has
already been factored out. 
Letting
\begin{equation}
\label{phad}
P_{\gamma p}^{had} = P_{\gamma}^{had}  \equiv P_{had} \ \ \ and \ \ \ 
 P_{\gamma \gamma}^{had} \approx   (P_{\gamma}^{had})^2
\end{equation}
The overlap function $A_{ab} (b)$ is 
\begin{equation}
\label{aob}
A_{ab}(b)={{1}\over{(2\pi)^2}}\int d^2\vec{q}{\cal F}_a(q) {\cal F}_b(q) 
e^{i\vec{q}\cdot \vec{b}}
\end{equation}
 where ${\cal F}$ is the Fourier transform of the b-distribution
of partons in the colliding particles. Normally, $A_{ab}$ is
 obtained using for ${\cal F}$ the electromagnetic form 
factors of the colliding hadrons. In general, for photons people have normally
used the form factor for a pion. We~\cite{liame} take a slightly different 
approach and calculate the `b- distribution' of the partons by taking the 
Fourier transform of the transverse momentum distribution of the partons, which
in the case of the photons is expected to be, at least for the perturbative 
part, 
\be
f(\kt) = {C \over (k_T^2 + k_{T,0}^2)}.
\ee
 The $k_{T,0}$ has actually been measured by ZEUS~\cite{ZEUS} to be 
$ 0.66 \pm 0.22$ GeV. It turns out that the form  of $A_{\gamma \gamma}$
with this transverse momentum ansatz and that for the pion form factor
ansatz, are the same, differing only in the value of the parameter 
$k_{T,0}$ which is $0.735$ GeV for the $\pi$ form factor case. Thus one can 
asses the effect of changing the ansatz for the $A_{ab}$ for photons by simply
changing the value of $k_{T,0}$.

For the soft part of the cross-section we use a parametrisation, 
\begin{equation}
\label{soft}
\sigma^{soft}_{\gamma p} =\sigma^0 +
{{A}\over{\sqrt{s}}}+{{B}\over{s}}
\end{equation}
we then calculate values for $\sigma^0, A$ and $B$ from a best fit 
\cite{thesis} to the low energy photoproduction data, 
starting with the Quark Parton Model (QPM) ansatz
$\sigma^0_{\gamma p}\approx {{2}\over{3}}\sigma^0_{pp}$ and the form 
factor ansatz for the $A_{\gamma p}$. The best fit value for \ptmin\ 
that we get is 2 GeV. It might be  possible to improve quality of the 
fit by using a energy dependent \Phad , but it needs to be investigted further.
The value of 2 GeV is also comparable to the value 1.6 GeV obtained~\cite{tjs}
from a fit to the description of minimum bias events in $pp/ \bar p p $ 
collisions.

For $\gamma \gamma$ collisions, we  repeat the QPM suggestion and propose
\begin{equation}
\sigma^{soft}_{\gamma \gamma}={{2}\over{3}} \sigma^{soft}_{\gamma p}.
\end{equation}
We now apply the  criteria and parameter set used in
 $\gamma p$ collisions to the case of photon-photon collisions,
i.e. $P_{\gamma}^{had}=1/204$, $\ptmin = 2$ GeV, 
A(b) from  the transverse momentum ansatz with the value $k_{T,0}=0.66$ 
GeV. The results of our calculation are shown 
\begin{figure}[hbt]
\leavevmode
\begin{center}
\mbox{\epsfig{file=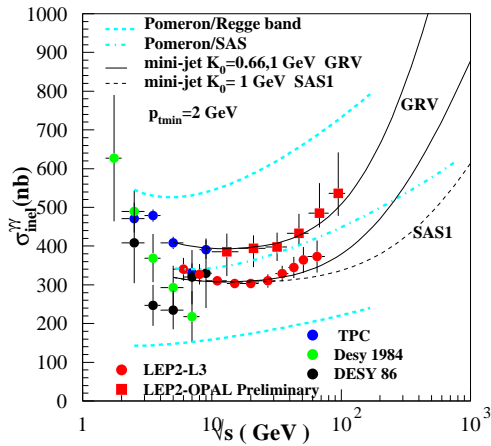,height=6cm}}
\end{center}
\caption{Total inelastic photon-photon cross-section in the eikonalized
 mini-jet model with  $\ptmin = 2\ GeV$,
 compared with data and Regge/Pomeron parametrization (see  Ref.~[5] 
 for details). The two lower mini-jet curves correspond to $k_{T,0}=1$ GeV 
 with GRV and SAS1 densities. The highest one is for GRV densities and 
 $k_{T,0}=0.66$ GeV.}
\label{Gamgam}
\end{figure}
Fig.~\ref{Gamgam}. 
The highest of the two full  lines corresponds exactly to the
 same parameter set used in the photoproduction case
and appears to be in good agreement with the preliminary results from the
OPAL~\cite{OPAL} Collaboration, whereas the L3 results, everything else being
the same, would favour a higher $k_{T,0}$ value.

The following can be noticed here from the  newer data and 
model calculation:
\begin{itemize}
\item The data on \siginel\ rise faster with the energy than the predictions of
the Regge-Pomeron ideas and  the rise is consistent with predictions of the 
eikonalised minijet models where the parameters are fixed by fits to the
photoproduction case.

\item  These calculations use also more uptodate photonic parton densities
and the parameters of the eikonalised minijet model are related here to measured
properties of photon induced reactions.
\end{itemize}

\section{Conclusions}
We can summarise the discussion so far as follows:
\begin{itemize}
\item The knowledge of parton content of the photon has improved
considerably in the last few years.
\item A combination of beamstrahlung and `strong' interactions   
of the photon can induce hadronic backgrounds at the future
\eplem\ and \gamgam\ colliders. The cross-section 
$\sigma (\gamma \gamma \rightarrow {\rm jets})\big|_{\ptmin}^{\sqrt s}$
of Eq.~\ref{Sigjet} is a good measure of the `messiness' that this can
cause.  From the calculation of total inelastic cross-section in the
eikonalised minijet model, for photoproduction as well as \gamgam\ 
collisions, 2 GeV seems to be good value for \ptmin. However,
due to the large number of other parameters involved, this can not be
taken to be a determination of \ptmin. A MC analysis of the newer
\twophoton\  data might help settle this further. {\bf The parametrisations
of Eqs.~(\ref{grv},\ref{eq:sas}), can be used conveniently to esitmate 
this `figure of merit' for a given collider by folding in with the photon
spectra.}

\item It seems that for $\sqrt s \leq 2$ TeV, for the current machine
designs,  the \eplem\ colliders can be free of `minijet' induced 
backgrounds. It would be preferable to repeat the analysis of Ref.~[4],
with the newer MC for photon induced processes that are available now.
An analysis of the energy flow due to the hadrons produced by the
interactions of the photons, should use the newer information
on the \siginel\ that has become available now.
\item For \gamgam\ colliders further analysis is necessary. The better
knowledge of  photonic parton densities available today as well as the
newer information on \siginel\ should be used as an input to the newer
analyses. At higher energies the saturation effects of the photon structure
function should also be included.
\end{itemize}

\end{document}